\documentclass[3p,times]{elsarticle}

\usepackage{ecrc}
\usepackage{slashed}
\usepackage{hyperref}

\usepackage{epstopdf}

\volume{00}

\firstpage{1}

\journalname{Nuclear Physics A}

\runauth{Ghiglieri}

\jid{nupha}

\jnltitlelogo{Nuclear Physics A}

\usepackage{graphicx}
\usepackage{amsmath,amssymb}

\def\sss{\scriptscriptstyle}

\def \bx {\mathbf{x}}

\def \nc {N_c}

\def \nf {N_f}

\newcommand{\Tint}[1]{{\hbox{$\sum$}\!\!\!\!\!\!\!\int\,}_{\!\!\!\!\raise-0.9ex\hbox{$\scriptstyle{#1}$}}}

\def\siml{{\ \lower-1.2pt\vbox{\hbox{\rlap{$<$}\lower6pt\vbox{\hbox{$\sim$}}}}\ }}
\def\simg{{\ \lower-1.2pt\vbox{\hbox{\rlap{$>$}\lower6pt\vbox{\hbox{$\sim$}}}}\ }}

\def \bx {\mathbf{x}}

\def \als {\alpha_{\mathrm{s}}}

\def \m2   {\mu^{2 \epsilon}}

\def\siml{{\ \lower-1.2pt\vbox{\hbox{\rlap{$<$}\lower6pt\vbox{\hbox{$\sim$}}}}\ }}
\def\simg{{\ \lower-1.2pt\vbox{\hbox{\rlap{$>$}\lower6pt\vbox{\hbox{$\sim$}}}}\ }}

\def\dgk{\frac{d\Gamma_\gamma}{d^3k}\bigg\vert}

\def\2to2{{2\leftrightarrow 2}}

\def\LO{{\mathrm{LO}}}
\def\NLO{{\mathrm{NLO}}}

\def\qhat {\hat{q}}

\def\k{{\bf{k}}}

\def\x{{\bf{x}}}
\def\OO{{\mathcal{O}}}

\def\nfd{n_{\!\sss F}}

\begin{document}

\begin{frontmatter}



\title{Theoretical aspects of photon production in high energy nuclear collisions}

\author{Jacopo Ghiglieri}

\address{McGill University, Department of Physics,\\3600 rue University, Montreal QC H3A 2T8, Canada}



\begin{abstract}
A brief overview of the calculation of photon and dilepton production rates
in a deconfined quark-gluon plasma is presented. We review leading order rates
as well as recent NLO determinations and non-equilibrium corrections. Furthermore,
the difficulties in a non-perturbative lattice determination are summarized.
\end{abstract}


\end{frontmatter}



\section{Introduction}
\label{intro}

Photons and dileptons have long been considered a key hard probe of the medium produced in high
energy heavy-ion collisions.  A chief advantage is that their coupling to the
plasma is weak, which means that reinteractions (absorption, rescattering)
of electromagnetic probes are expected to be negligible. These probes hence 
carry direct information about their formation process to the detectors,
unmodified by hadronization or other late time physics. 


Experimentally, there are now detailed data on both real photon and dilepton production alike
 at RHIC 
 and the LHC. 
An overview of experimental data was presented at this conference in \cite{Lijuantalk}.
Photons and dileptons arising
from meson decays following hadronization are subtracted from the data experimentally or form
part of the ``cocktail''; 
theoretically one then needs to deal with several sources during the evolution of the fireball:
\emph{``prompt'' } photons and dileptons, produced in the scattering
of partons in the colliding nuclei, 
\emph{jet photons}, arising from the interactions and fragmentations of jets,
\emph{thermal}  photons and dileptons, produced by interactions of the (nearly) thermal constituents of 
the plasma and 
\emph{hadron gas}  photons and dileptons, produced in later stages. Phenomenologically,
one then needs to convolute \emph{microscopic production rates} over the \emph{macroscopic
spacetime evolution} of the medium produced in the collision, which is governed by an effective
hydrodynamical description. A review on the application of hydrodynamics to heavy-ion collisions
has been presented in \cite{Niemitalk}.

In this contribution we will concentrate on an overview and on some recent results
on the microscopic rates and their computation, focusing on the thermal phase. 
We will treat real photons and virtual photons (dileptons). An overview of
the phenomenological aspects has been presented in \cite{Elenatalk}.


From a theorist's perspective, the determination of the photon
and dilepton rates boils down to evaluating the same function with different
kinematical conditions and prefactors. In more detail, 
at leading order in QED (in $\alpha$) and to all orders in QCD
the photon production rate per unit phase space is
(see for instance Ref.~\cite{LeBellac})
\begin{equation}
\label{defratephoton}
\frac{dN_{\gamma}}{d^4 X d^3 \k} \equiv \frac{d\Gamma_{\gamma}}{d^3 \k}
= - \frac{1}{(2\pi)^3 2 |\k|}
W^<(k^0=k), 
\end{equation}
whereas the dilepton rate reads
\begin{equation}
\label{defratedilepton}
\frac{d\Gamma_{l\bar l}}{d^4K}=-\frac{2\alpha}{3(2\pi)^4K^2}	
W^<(K) \, \theta( (k^0)^2 - \k^2) \,.
\end{equation}
where $K^2=(k^0)^2-k^2$ is the virtuality of the dilepton pair,
assumed much greater than $4m_l^2$. Both rates are given in 
terms of the photon polarization $W^<(K)$, which reads
\begin{equation}
\label{defPi}
W^<(K) \equiv \int d^4X e^{i K\cdot X}
   \mathrm{Tr}\rho J^\mu(0)J_\mu(X)\,.
\end{equation}
The main elements for the determination of this expression are
the \emph{electromagnetic current} $J$, which describes the coupling
of photons to the medium degrees of freedom and the \emph{density operator}
$\rho$, determining the state of these d.o.f.s. In most cases the equilibrium
approximation is taken, so that $\rho\propto \exp(-\beta H)$ and Eq.~\eqref{defPi}
becomes a thermal average. When convoluting
over the macroscopic evolution this corresponds to assuming a local equilibrium
in each discrete medium element.\footnote{Eq.~\eqref{defPi} is strictly speaking
not correct in a full out-of-equilibrium case, as translation invariance has been used
to factor out one spacetime integration. The generalization is however straightforward.}
Later on we will mention some calculations that go beyond the local equilibrium approximation.
Finally, an action or Lagrangian is necessary for the evaluation of Eq.~\eqref{defPi}, describing
the propagation and interaction of the medium d.o.f.s.

Approaches for the evaluation of Eq.~\eqref{defPi} in the thermal phase include
perturbation theory, the lattice and holographic techniques. In the first, 
the standard QCD action is employed and the current is simply the Dirac current. 
The thermal average can also be generalized to out-of-equilibrium situations. This method
is justified when $g\ll 1$; when applied to a more realistic coupling of $\als\sim 0.3$
it becomes an extrapolation, whose systematics may not be in full control. In Secs.~\ref{sec_lo}
and \ref{sec_nlo}
we will describe the basics of perturbative calculations and show how recent next-to-leading
order determinations can be used to test the reliability of the leading-order computations,
which are widely employed in phenomenological descriptions.

On the lattice, one employs the Euclidean QCD action and computes the path integral numerically 
in the equilibrium case. 
Eq.~\eqref{defPi} is however defined in Minkowskian space. Hence, a highly nontrivial analytical
continuation is required to approach the ``real world''. We will illustrate the current status
in Sec.~\ref{sec_lattice}.

Holographic techniques draw from the AdS/CFT correspondence \cite{Maldacena:1997re,Witten:1998qj,Gubser:1998bc}. The $\mathcal{N}=4$
supersymmetric Yang-Mills action is employed and coupled to a $U(1)$ field mimicking the photon's.
Holography is used for the evaluation at strong coupling, whereas perturbation theory can be employed
at weak coupling, giving the opportunity of studying the transition between the two regimes within the same
theory. In approaching the real world, the challenge lies in the extrapolation towards $\mathcal{N}=0$, 
$\nc=3$ QCD. Leading order calculations for photon and dilepton production can be found in \cite{CaronHuot:2006te} , $\OO(1/\lambda)$
corrections have been computed in \cite{Hassanain:2012uj} and generalization to an out-of-equilibrium, thermalizing medium can be found 
in \cite{Baier:2012tc,Baier:2012ax,Steineder:2013ana}.

This contribution is organized as follows: in Sec.~\ref{sec_lo} we describe in some detail
the leading-order calculation of the photon rate, which we use as an example. In the following
section~\ref{sec_nlo}, we show NLO calculations and non-equilibrium generalizations. In Sec.~\ref{sec_lattice}
we describe the non-perturbative approach and finally in Sec.~\ref{sec_summary} we draw our conclusions.

\section{Perturbation theory at LO: the photon example}
\label{sec_lo}
\begin{figure}[ht]
\begin{center}
\includegraphics[width=8cm]{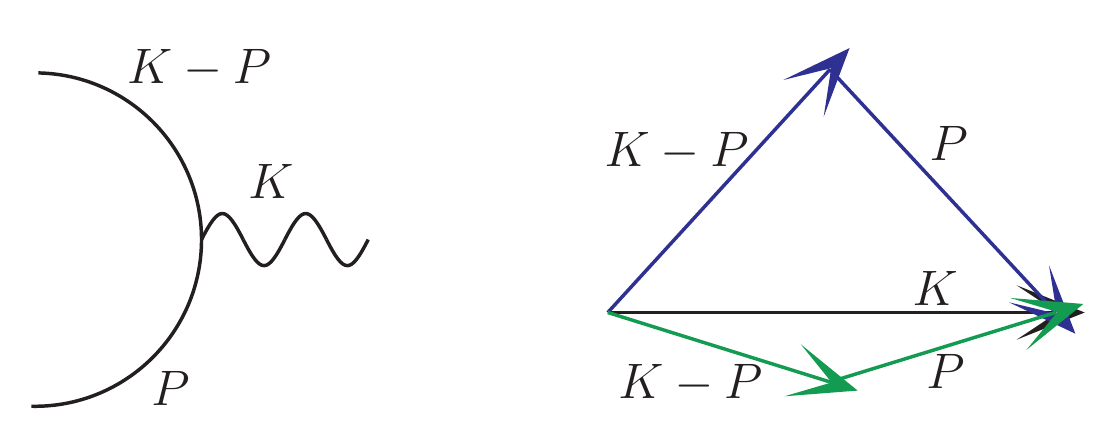}	
\end{center}
\vspace{-0.5cm}
\caption{Left: momentum assignments at one of the two current insertions. Right: the 2$\leftrightarrow$2  (blue) and collinear (green) processes
 are schematically represented.}
\label{fig_moms}
\end{figure}

\begin{figure}[ht]
	\begin{center}
		\includegraphics[width=10cm]{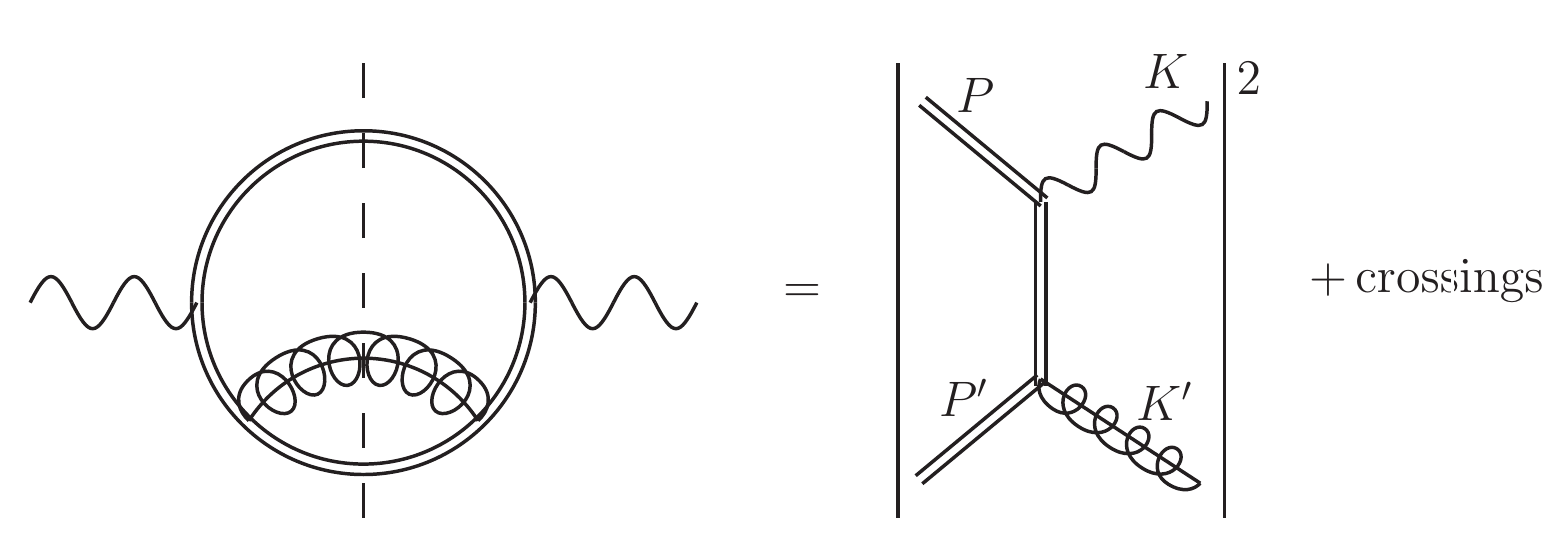}
	\end{center}
	\vspace{-0.5cm}
	\caption{The two-loop diagram on the left corresponds to the square of the amplitude 
	of the diagram on the right and on the squares of its crossing. The interference terms 
	arise from the two-loop diagram where the gluon is exchanged between the two fermionic 
	lines. Double internal lines stand for hard particles, \emph{i.e.}, particles whose momentum is 
	$\OO(T)$ in at least one component. Plot taken from \cite{Ghiglieri:2013gia}.}
	\label{fig_22}
\end{figure}

At the lowest order in pertubation theory Eq.~\eqref{defPi} vanishes for $k^0=k$, as the two bare Wightman
propagators in the simple quark loop it generates cannot be simultaneously put on shell. In
other words, on-shell quarks cannot radiate a physical photon. It is then necessary to
kick at least one of the quarks off-shell. This implies that the leading order is 
$\OO(\alpha g^2)$. Furthermore, the different processes contributing to the rate,
corresponding to different kinematical regions
are most naturally classified based on the scaling and virtuality. If $K$ is the photon momentum,
let us call $P$ and $K-P$ the fermion momenta at a current insertion in Eq.~\eqref{defPi}. Then,
as Fig.~\ref{fig_moms} schematically shows,
there are two distinct ways at LO to satisfy momentum conservation. $2\leftrightarrow 2$ processes correspond to the 
Compton-like and annihilation processes arising from the cuts of the two-loop diagrams obtained by adding an extra gluon, such as the one shown in Fig.~\ref{fig_22}.
Their evaluation boils down schematically to a kinetic-theory picture: an integral over the (Lorentz-invariant) phase space of the
matrix elements squared for the processes, folded over the statistical functions $f$ of the medium constituents, i.e.
\begin{equation}
\frac{d\Gamma_\gamma}{d^3k}\bigg\vert_{2\leftrightarrow 2}\sim \int_\mathrm{phase\, space}f(p)f(p')(1\pm f(k'))\,\left\vert \mathcal{M}
\right\vert^2\delta^4(P+P'-K-K')\,.
\label{schematic}
\end{equation}
The integration over the bare matrix elements encounters the IR divergence of massless Compton scattering. At finite
temperature, it is removed by collective plasma excitations. This is implemented by resumming the Hard Thermal Loop (HTL)
self-energy \cite{Braaten:1989mz} in the intermediate quark, rendering the $2\leftrightarrow 2$ processes finite
and non-analytic in $g$ through a $\log(g)$ term \cite{Kapusta:1991qp,Baier:1991em}.

\begin{figure}[ht]
	\begin{center}
	\begin{minipage}{0.08\textwidth}
		$\dgk_\mathrm{coll}$=
		\end{minipage}
		\begin{minipage}{0.75\textwidth}
		\includegraphics[width=13.2cm]{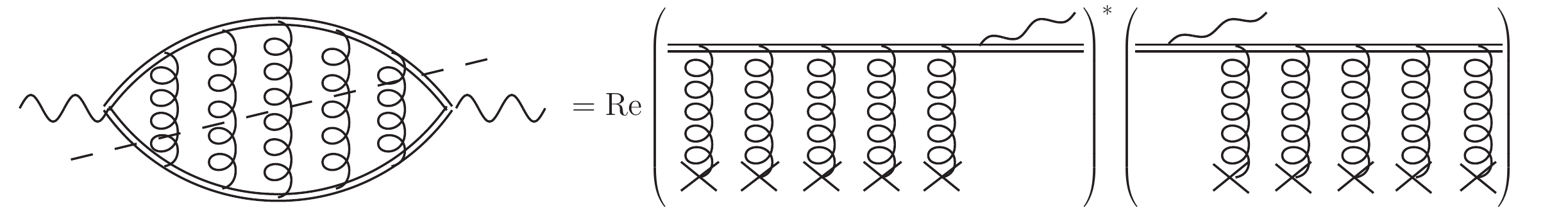}
		\end{minipage}
	\end{center}
	\caption{The  ladder diagrams that need to be resummed to account 
	for the LPM effect in the collinear region. The cut shown here corresponds to 
	the interference term on the right-hand side. The rungs on the l.h.s. are HTL 
	gluons in the Landau cut. On the r.h.s., the crosses at the 
	lower end of the gluons represent the hard scattering centers, either 
	gluons or fermions. Plot taken from \cite{Ghiglieri:2013gia}.}
	\label{fig_lpm}
\end{figure}

\emph{Collinear} processes, first introduced in \cite{Aurenche:1998nw}, contribute at LO. A soft collision
with the medium has an order-$\alpha$ chance of inducing the radiation of a collinear photon.
Its long formation time can be of the same order of the soft scattering rate, thus causing multiple
such scatterings to contribute at the same order and possibly interfere, in what is called the Landau-Pomeranchuk-Migdal
(LPM) effect. Its treatment requires the resummation of an infinite number of ladder diagrams \cite{Arnold:2001ba,Arnold:2001ms},
 as shown in Fig.~\ref{fig_lpm}.

\section{Beyond the LO photon rate: improvements and extensions of pQCD }
\label{sec_nlo}
In this Section we will give an overview of recent results which allow to extend
and stretch the pQCD calculation just illustrated. 
As we mentioned in the introduction, perturbation theory at finite temperature, with its thermal scales
and needs for resummations, is properly defined when $g\ll 1$. A next-to-leading order calculation
can then be of great help in establishing the reliability of the leading-order rates, which are
widely employed in phenomenology. In Sec.~\ref{sub_nlo_gamma}
we will review the recent NLO order calculation of the photon rate \cite{Ghiglieri:2013gia},
whereas in Sec.~\ref{sub_dilepton} we will present a recent NLO calculation for dileptons
\cite{Laine:2013vma}. In Sec.~\ref{sub_jf} we will discuss how the LO rates can be generalized
to include non-equilibrium viscous effects. Finally, a matrix-model inspired parametrization 
of the statistical functions close to the critical temperature has been used to compute
photon and dilepton rates close to $T_c$ and was presented in \cite{Shulintalk}.

\subsection{Beyond leading order}
\label{sub_nlo_gamma}
It is well known in perturbative thermal field theory that loop corrections involving
soft bosons, in our case gluons, are suppressed by a factor of $g$ only. Hence, the
NLO correction to the photon rate is just an $\OO(g)$ correction and schematically
it arises from all regions where existing gluons can become soft or extra soft
gluons can be added.

One such place is the collinear region. An $\OO(g)$ correction is given by
the NLO contribution to the soft scattering rate and to the jet quenching
parameter $\qhat$, as computed by Caron-Huot in \cite{CaronHuot:2008ni}. In
that paper, analyticity properties of retarded amplitudes are used to map
the intricate real-time (Minkowskian) four-dimensional calculation of the
soft contributions to a much simpler Euclidean three-dimensional problem, 
basically extending the applicability domain of dimensional reduction from
time-independent quantities to the lightcone. This is at the base
of the recent lattice studies of $\qhat$ \cite{Panero:2013pla,Laine:2013lia},
presented in \cite{Marcotalk}.
In terms of our calculation, this corresponds to considering the effect
of a one-loop rung amid the tree-level ladder of Fig.~\ref{fig_lpm}, and similarly
of considering the $\OO(g)$ shift in the thermal mass of the collinear quarks \cite{CaronHuot:2008uw}.\footnote{
Recently \cite{Brandt:2014uda,Harveytalk} it has been shown that a class of screening masses can be obtained from a differential
equation where the potential is the transverse scattering kernel. The non-perturbatively
determined one seems to give the best agreeement with the lattice-calculated screening masses.}

A new set of processes arises at the next order in the collinear expansion. If 
at leading order the angle has to be of order $g$, at NLO it can be extended
to $\sqrt{g}$, giving rise to semi-collinear processes, where, besides
the usual soft spacelike scattering, a soft timelike gluon (a plasmon) can  also
be absorbed and induce radiation. The evaluation of this region requires a modified
version of $\qhat$ keeping track of these features. Its relevance for jet energy loss
at higher order is being investigated \cite{jets}.


Finally, we mentioned how at leading order the region where the intermediate quark 
is soft needs to be treated with care, requiring HTL resummation. At NLO one has
the possibility of adding soft gluons to an already soft quarks, giving rise to
intricate amplitudes with several HTL propagators and vertices. However, at equilibrium
these expressions can be written in terms of retarded functions only and the aforementioned
analyticity properties allow a deformation of the integration contour away from the real
axis, where the intricacies of HTLs lie, to a region where the amplitudes are much simpler,
leading to a compact expression. 

\begin{figure}
	\includegraphics[width=0.495\textwidth]{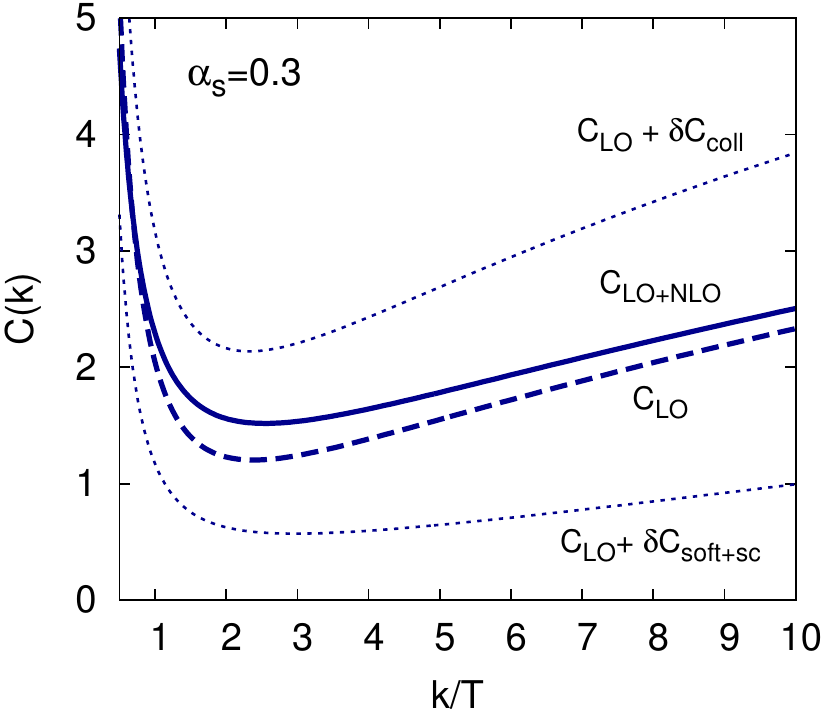}
	\includegraphics[width=0.495\textwidth]{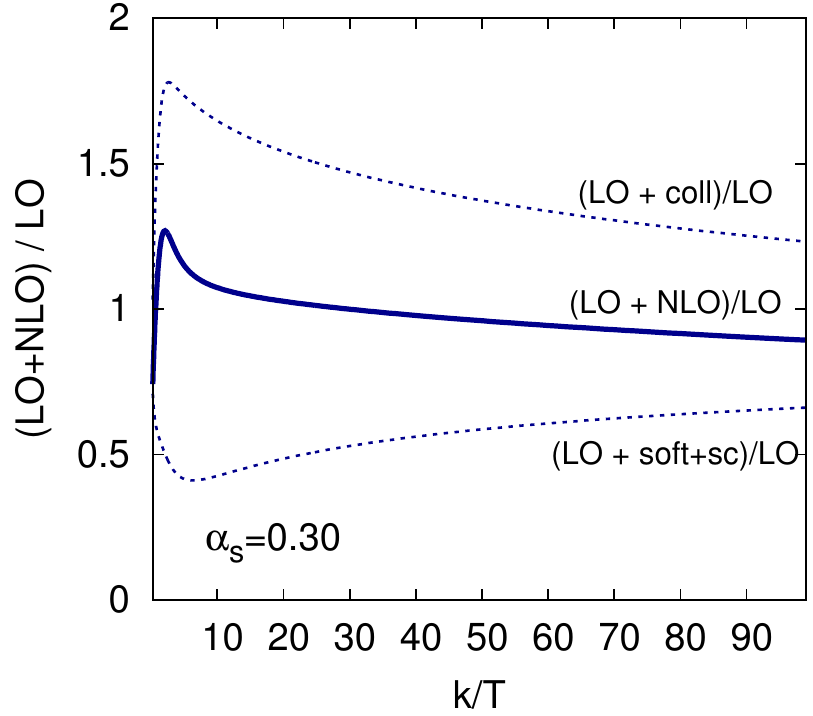}
   \caption{Left: the function, $C(k/T)\equiv d\Gamma/(d^3k)(4 \alpha_{\rm EM} \nfd(k) g^2 T^2/(3 k))^{-1} $, parametrizing the photon emission rate 
      for $\nc=\nf=3$ and $\als=0.3$.
       The full next-to-leading order function
         ($C_{\LO+\NLO}$) is a sum of the leading-order result ($C_{\LO}$), a 
         collinear correction ($\delta C_{\rm coll}$),  and a
         soft+semi-collinear correction ($\delta C_{\rm soft+sc}$).  
         The dashed curve  labeled
         $C_{\LO}+\delta C_{\rm coll}$ 
         shows the result when  only the collinear correction
         is included, with the analogous notation  for the  $C_{\LO} + \delta C_{\rm soft+sc}$ curve.  
		 The difference between the dashed curves provides a 
         uncertainty estimate  for the NLO calculation.
      Right: the ratio between the LO+NLO rate over the LO rate for the same parameters. Plot taken from \cite{Ghiglieri:2013gia}.}
	\label{fig_plot}
\end{figure}

In Fig.~\ref{fig_plot} we plot the rate (in units of the leading-log coefficient) and the ratio of LO+NLO correction over LO rate 
for $\nc=\nf=3$ and $\als=0.3$. Both plots show how in the phenomenologically relevant region the NLO correction represents an 
$\OO(20\%)$ increase which comes about from a cancellation between a much larger positive correction from the collinear sector
and a negative one from the semi-collinear and soft regions.
This cancellation appears largely accidental, confirmed by the fact that at larger momenta the negative contribution
overcomes the positive one. Hence, the band formed by these two curves can be taken as an uncertainty estimate of
the LO calculation and should be considered in phenomenological analyses.

\subsection{Dileptons}
\label{sub_dilepton}
\begin{figure}[ht]
\begin{center}
\includegraphics[width=14cm]{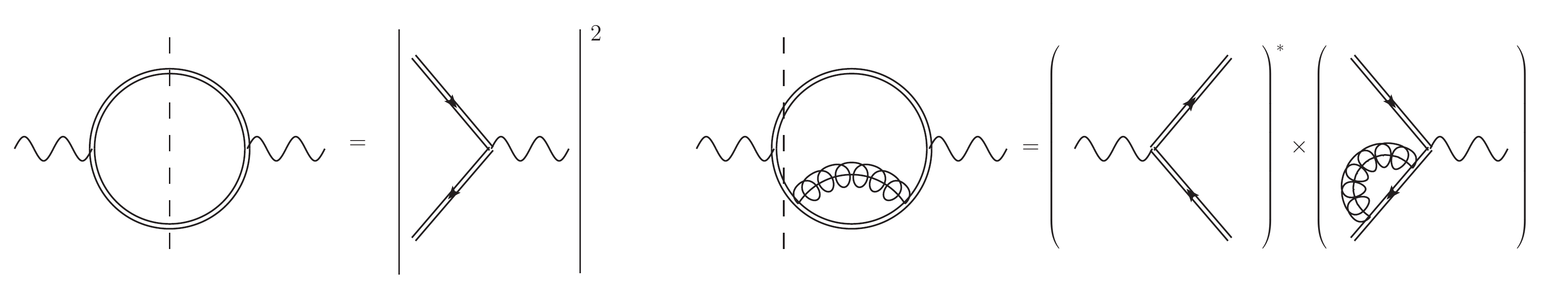}
\end{center}
\caption{Left: the Born diagram and its cut, corresponding to the thermal Drell-Yan process. Right: a two-loop
correction with a cut corresponding to a virtual correction to the Drell-Yan process.}
\label{fig_dilepton}	
\end{figure}

The dilepton invariant mass corresponds to the virtuality of the photon in Eq.~\eqref{defPi}, $K^2>0$.
This finite $K^2$ now allows a Born term at leading order in the loop expansion, which corresponds to
a \emph{thermal Drell-Yan process}, i.e. the annihilation of a quark-antiquark pair into a virtual
photon, as shown on the left of Fig~\ref{fig_dilepton}.  At the next order in the loop expansion one can have both real and virtual corrections
to this process. In the former case  they correspond to $2\leftrightarrow 2$ processes, as shown in 
Fig.~\ref{fig_22}, whereas the latter case is shown on the right in Fig.~\ref{fig_dilepton}. Cancellations of
 soft and collinear divergences are at play between these corrections. 
 However, if the frequency and/or the virtuality
 of the photon become soft, HTL and/or collinear resummations become necessary \cite{Braaten:1990wp,Aurenche:2002wq},
  along the lines of what has been described for the real photon case.

\begin{figure}[ht]
\begin{center}
\includegraphics[width=8cm]{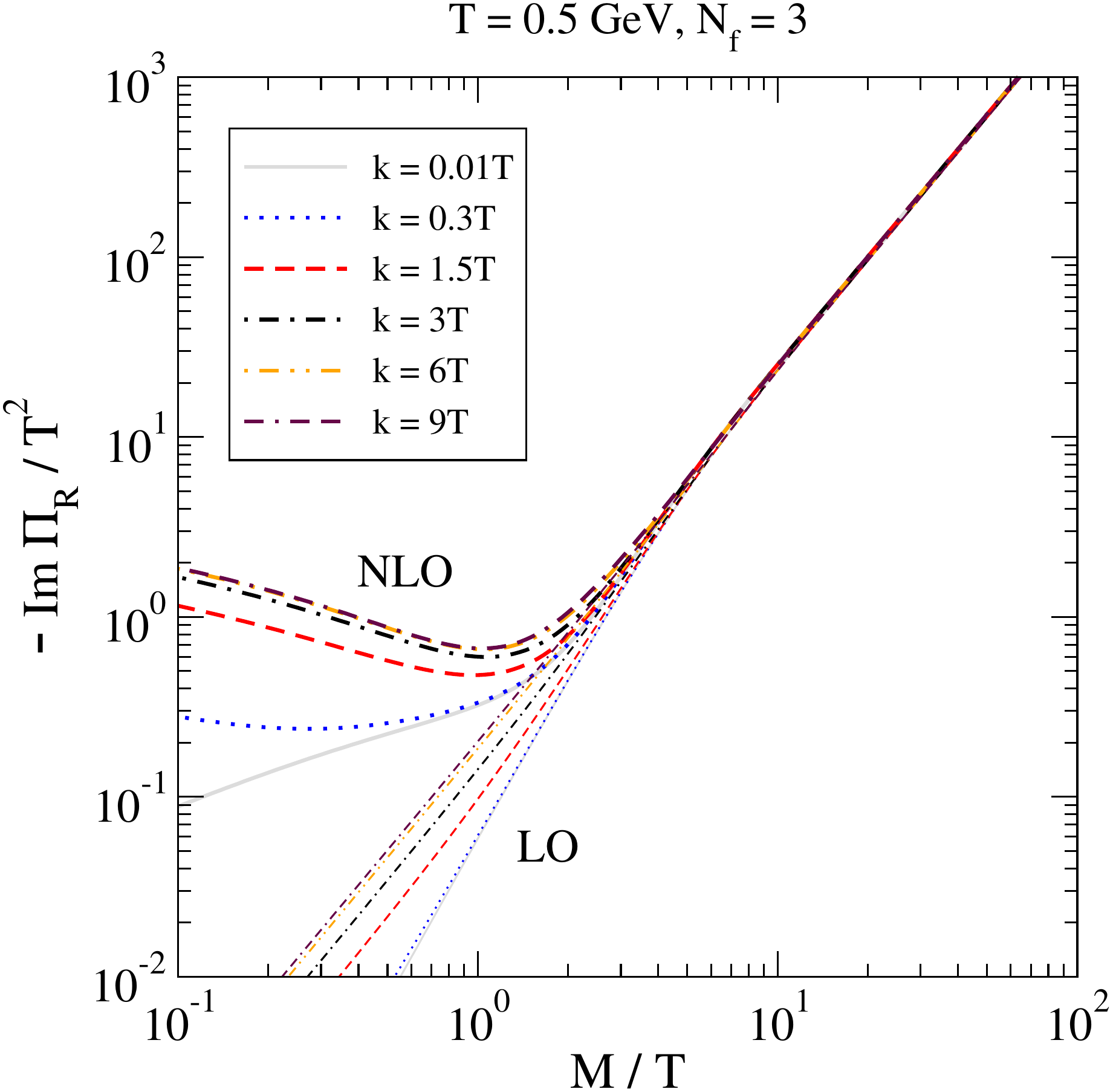}
\includegraphics[width=8cm]{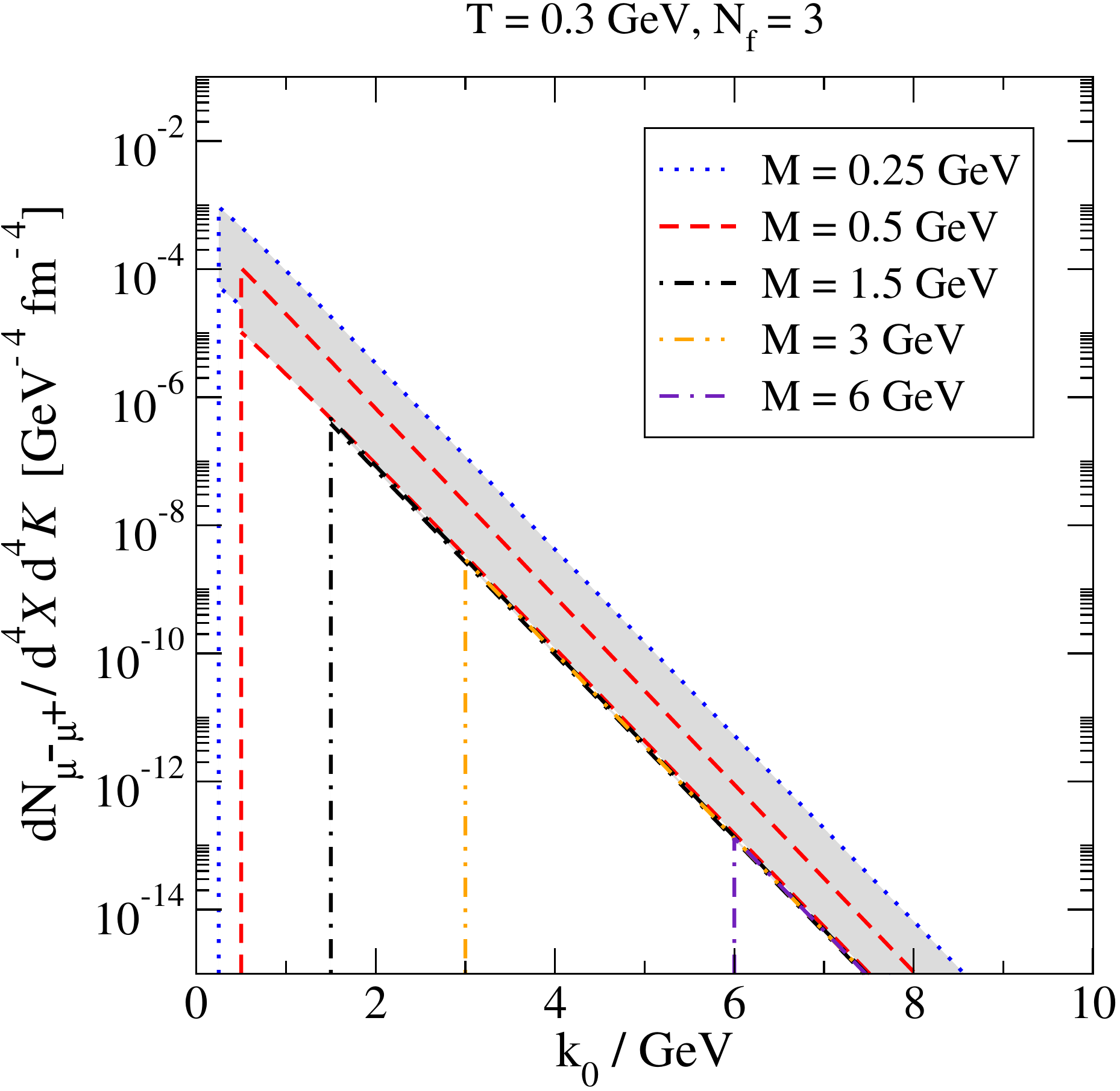}	
\end{center}
\caption{Left: the imaginary part of the retarded polarization, corresponding to the spectral function
of the e/m current $J$. Right: the dilepton rate in physical units. Both plots are taken from \cite{Laine:2013vma}.}
\label{fig_mikko}
\end{figure}

Recently, a complete NLO calculation of the dilepton rate for $K^2\simg T^2$, $k>0$ has been completed in 
\cite{Laine:2013vma}.\footnote{LPM resummation has been added in \cite{Ghisoiu:2014mha}.} As the left pane in Fig.~\ref{fig_mikko} shows, the thermal Drell-Yan process (the LO)
receives increasingly larger corrections at diminishing $M=\sqrt{K^2}$, until the NLO correction overtakes
it, signalling the need for resummations. The right pane shows the rate in physical units. For the smallest masses
an error band is obtained by varying the renormalization scale for the running coupling by a factor of 2 in each
direction.
A NLO calculation of the dilepton rate for $K^2\sim (gT)^2$, with HTL and collinear resummations, is underway \cite{dileptons}.




\subsection{Beyond thermal equilibrium}
\label{sub_jf}
As we mentioned, the medium produced in heavy-ion collisions undergoes a viscous
hydrodynamic evolution. Assessing thus the importance of the deviation from equilibrium
is of great relevance for a quantitative description of photon flow data, such as the
``$v_2$ puzzle'' widely reported at this conference. To this end,
$2\leftrightarrow 2$ processes have been computed in an off-equilibrium setting \cite{Shen:2013cca}, 
as reported in \cite{Chuntalk}. The simple kinetic picture of Eq.~\eqref{schematic} can be readily
generalized, by replacing the equilibrium distributions $f$ with first-order perturbed ones which
depend on the anisotropic tensor $\pi^{\mu\nu}$.\footnote{An altogether similar treatment is also
applied to the hadronic phase.} 
 In the soft region, equilibrium Hard Thermal Loops are replaced by 
off-equilibrium Hard Loops. We refer to \cite{Chuntalk} for the phenomenological implications; from 
a theoretical standpoint, we remark that it would be very interesting to see an anisotropic extension
of collinear processes, which could be more intricate due to the instabilities of gluonic
Hard Loops \cite{Romatschke:2003ms,Arnold:2003rq}.
\section{Lattice}
\label{sec_lattice}
As we mentioned in the introduction, a direct determination of Eq.~\eqref{defPi} is hampered 
by its Minkowskian nature. The corresponding Euclidean correlator, accessible by lattice 
methods, is
\begin{equation}
\label{defeuclid}
W_E(\tau,\k)\equiv \int d^3x \left\langle J_\mu(\tau,\bx)J_\mu(0)\right\rangle e^{i\k\cdot\bx}\,.
\end{equation}
This Euclidean correlator is related by analytical continuation to the Minkowskian one: $W_E(\tau,\k)=W^<(i\tau,\k)$,
which allows to write the former in terms of the electromagnetic spectral function as
\begin{equation}
\label{spf}
W_E(\tau,\k) =\int_0^\infty\frac{dk^0}{2\pi}\rho_J(k^0,\k)\frac{\cosh\left(k^0(\tau-1/(2T))\right)}{\sinh\left(\frac{k^0}{2T}\right)}.
\end{equation}
One is then faced with the numerical inversion of the r.h.s of Eq.~\eqref{spf}, starting from a discrete set of points with errorbars
on the l.h.s. At vanishing momentum ($k=0$) the inversion has been performed using Bayesian MEM techniques or fitting Ans\"atze (see 
\cite{Ding:2010ga,Ding:2013qw} for recent results). The $k=0$ spectral function encodes the physics of electric conductivity in the 
transport peak at the origin and the dilepton rate for finite $k^0$. The obtained result shows a similar trend to the resummed HTL
calculation of \cite{Braaten:1990wp} and at large values of the virtuality is well described by the perturbative Born term.

\begin{figure}
\begin{center}
\includegraphics[width=8cm]{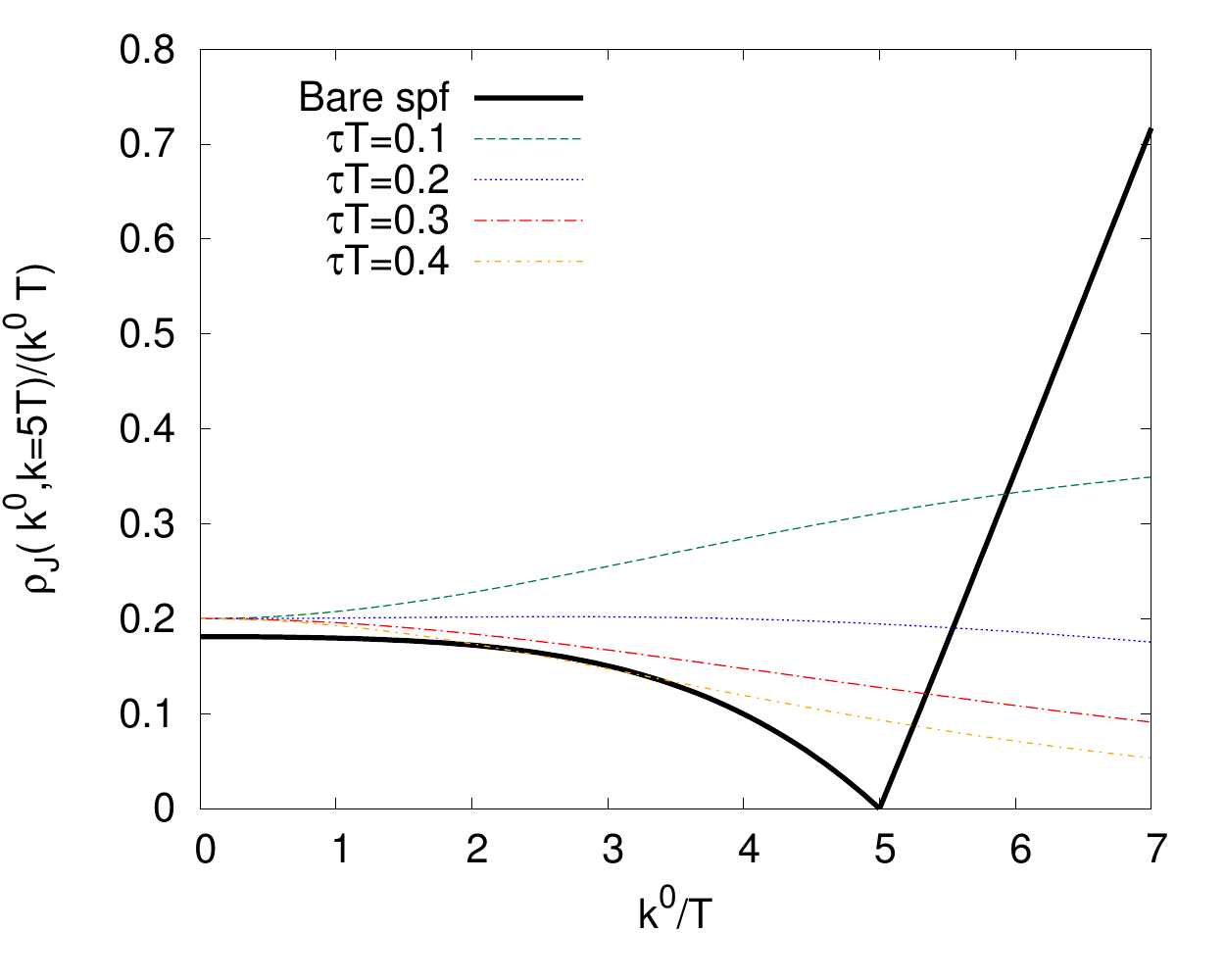}	
\end{center}	
\vspace{-0.5cm}
\caption{The Born spectral function ($g=0$) in black, for $k=5T$, divided by the frequency. The superimposed dashed lines are 
$k^0\,\cosh\left(k^0(\tau-1/(2T))\right)/\sinh\left(\frac{k^0}{2T}\right)$ for different values of $\tau$. }
\label{fig_spf}
\end{figure}

\begin{figure}
\begin{center}
\includegraphics[width=6cm]{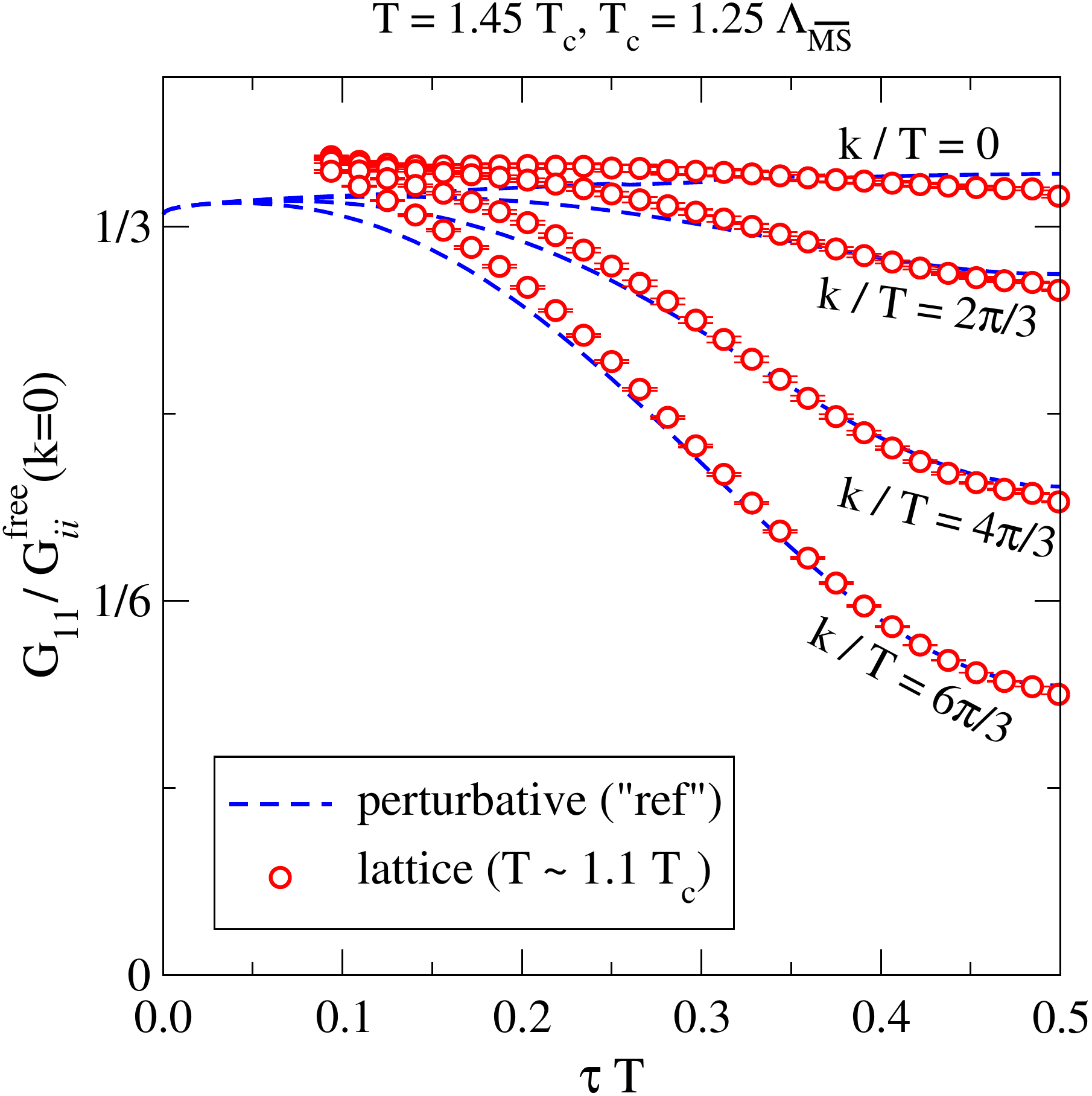}
\includegraphics[width=6cm]{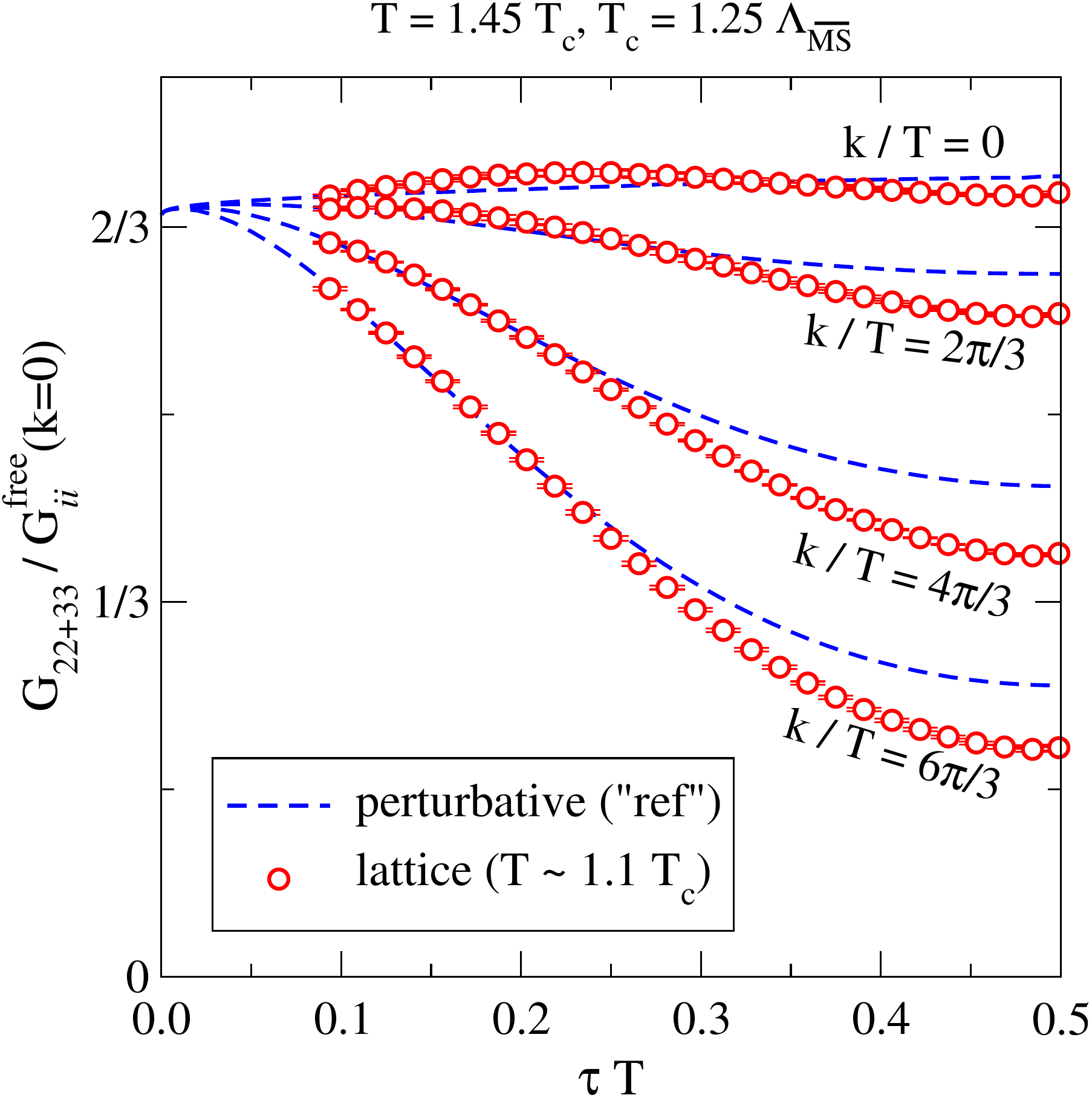}
\end{center}	
\caption{The left and right panes show the comparison of the perturbative and lattice data, as explained in the main text. Both 
are taken from \cite{Laine:2013vma}. }
\label{fig_spf_2}
\end{figure}

At non-vanishing momentum, $k>0$, the spectral function describes the physics of spacelike scattering with medium constituents ($0<k^0<k$),
real photon production ($k^0=k$) and dilepton production ($k^0>k$). Although a determination of $W_E$ for $k>0$ exists \cite{Ding:2013qw},
no attempt at spectral function reconstruction has been performed so far. It is however straightforward to use Eq.~\eqref{spf} directly, i.e. to obtain
$W_E$ from a perturbative determination of the spectral function, as carried out in \cite{Laine:2013vma}. The leading-order
spectral function (complemented with higher-order vacuum contributions) is shown to be in qualitative agreement with the lattice
data. It is to be remarked, however, that extremely interesting features, such as the real photon contribution $k^0=k$, vanish in the LO spectral function. 
In more detail, Fig.~\ref{fig_spf} shows the bare spectral function (without any higher-order vacuum contributions), 
normalized by the frequency, with superimposed
dashed lines for the ratio of hyperbolic functions in Eq.~\eqref{spf} (multiplied by frequency) for different values of $\tau$. 
It is the overlap of these curves that determines the Euclidean correlator and, as the figure shows, it is dominated by the
approximate linear rise of the dilepton ($k^0>k$) part of the spectral function. Fig.~\ref{fig_spf_2} shows the comparison \cite{Laine:2013vma} of
the perturbative \cite{Laine:2013vma} and lattice  calculations \cite{Ding:2013qw} in the longitudinal (left) and transverse (right) channels.  We
refer to \cite{Laine:2013vma} for a discussion on the \emph{caveats} of this comparison.
\section{Summary and conclusions}
\label{sec_summary}
The knowledge of the production rates and of the associated theoretical uncertainty
 is an essential
ingredient for the phenomenological description of electromagnetic hard probes. 
In this contribution
we have illustrated the basics of the calculation of the LO pQCD rates, widely employed in
phenomenology. These calculations often require the inclusion of physically distinct processes
and careful resummations of soft and collinear physics. The reliability of these LO rates
can be tested through NLO calculations.
These indeed provide a first estimate of the theory error budget for photons \cite{Ghiglieri:2013gia}
and large-mass dileptons \cite{Laine:2013vma}. As a byproduct, technological advancements have been 
obtained in the form of factorizations, sum rules for Hard Thermal Loop and calculations of spectral
functions \cite{Laine:2013vpa,Laine:2013lka} that are employable in other domains, such as jet propagation and quenching.

For what concerns the lattice, we have reviewed the difficulties in the extraction
of the spectral function. Recent results \cite{Ding:2010ga,Ding:2013qw} are available
for $k=0$ and a first comparison with perturbation theory was performed, for $k\ne 0$,
at the Euclidean correlator level \cite{Laine:2013vma}. Furthermore, important ingredients
for the perturbative calculations, such as the soft scattering rate, are now being
determined non-perturbatively on the lattice \cite{Panero:2013pla}, opening the way
for possible future  ``hybrid'' calculations, where non-perturbative inputs
are used in perturbative calculations.

\emph{Acknowledgements}
This work was supported by the Institute for Particle Physics (Canada) and the Natural
Sciences and Engineering Research Council (NSERC) of Canada. I would like to acknowledge
the Mainz Institute for Theoretical Physics (MITP) for enabling me to complete a portion
of this work.
\bibliographystyle{elsarticle-num}
\bibliography{photonbib.bib}



%
%
%

\end{document}